\documentclass{PoS}
\title{Effects Of Meson-nucleon Dynamics In A Relativistic Approach To Medium-mass Nuclei}

\ShortTitle{Effects Of Meson-nucleon Dynamics In A Relativistic Approach To Medium-mass Nuclei}

\author{\speaker{Elena Litvinova}\\
Department of Physics, Western Michigan University, Kalamazoo, MI 49008, USA\\
National Superconducting Cyclotron Laboratory, Michigan State University,\\
East Lansing, MI 48824, USA      \\
      E-mail: \email{elena.litvinova@wmich.edu}}
\author{Caroline Robin\thanks{US-NSF PHY-1404343.}\\
Department of Physics, Western Michigan University, Kalamazoo, MI 49008, USA      \\
      E-mail: \email{caroline.robin@wmich.edu}}

\abstract{Recent developments in the relativistic nuclear field theory (RNFT) are reviewed. Based on the covariant meson-nucleon Lagrangian of quantum hadrodynamics, RNFT connects consistently the high-energy scale of heavy mesons, medium-energy range of pion and the low-energy domain of emergent collective vibrations (phonons) in a parameter-free way. Mesons and phonons build up the effective interaction in various channels, in particular, the phonon-exchange part takes care of the retardation effects, which are of great importance for the fragmentation of single-particle states, spreading of collective giant resonances and soft modes, quenching and beta-decay rates with significant consequences for astrophysics and for the theory of weak processes in nuclei. As examples, the recently discovered impact of isospin dynamics on the nuclear shell structure and the isospin-flip pairing vibrations are discussed.      

}

\FullConference{The 26th International Nuclear Physics Conference\\
		 11-16 September, 2016\\
		 Adelaide, Australia}

\begin{document}

Understanding the precise structure of atomic nuclei is one of the major problems of research in natural sciences. As a mesoscopic system, the atomic nucleus provides a connection between the macroscopic world with pronounced statistical regularities and the microscopic world, where individual quantum states form discrete patterns underlying large-scale physics. For instance, modeling the elemental composition of such macroscopic objects as the solar system, galaxies and stars is largely based on the knowledge about the intrinsic properties of atomic nuclei. 
%On the cosmic scale of astrophysical objects, the uncertainties in the nuclear physics input cause tremendous error propagation, which can render the %astrophysical modeling extremely unreliable.

In the context of global modeling of the nuclear landscape, it is essential to build a high-precision solution of the nuclear many-body problem, which enables a consistent computation of masses, matter and charge distributions, spectra, decay and various reaction rates within the same framework at zero and finite temperatures throughout the entire nuclear chart. Such a modeling is a cornerstone for modern studies of r-process nucleosynthesis,  leading to the formation of heavy elements in the universe.
%, which is regarded to take place in catastrophic events, such as type II (core collapse) supernovae or neutron star mergers. 
These studies are not only very sensitive to the accuracy of the nuclear physics input, but also require detailed and comprehensive information about nuclei, which can not be synthesized in laboratories. Thus, our understanding of chemical evolution and elemental composition of the universe relies on a fundamental, accurate and predictive nuclear structure theory. 
%In order to have a reliable predictive power, theoretical approaches should be as fundamental and universal as possible. 
However, in spite of many advances made over the past decades of research, a global high-precision theory for the description of structure properties of nuclei is still very far from completeness.

Although the constituents of nuclei, protons and neutrons, are composite particles with complex internal structure, describing the nuclei without resolving nucleonic internal degrees of freedom remains the prevailing paradigm in the modern low-energy nuclear physics. The underlying theory of protons and neutrons is Quantum Chromodynamics (QCD), which has lately advanced to the description of light nuclei on the lattice \cite{BeaneChangCohenEtAl2012,YamazakiIshikawaKuramashiEtAl2015,ChangDetmoldOrginosEtAl2015}. While direct extensions of the Lattice QCD (LQCD) calculations to medium mass and heavy nuclear systems are not yet possible technically, LQCD is capable of providing effective nucleon-nucleon potentials, which can be used in combination with many-body microscopic methods (to be distinguished from those using collective coordinates) based on the nucleonic degrees of freedom: (i) "ab-initio" approaches, (ii) configuration interaction (CI) models (known also as shell-models) and (iii) density functional theories (DFT).

\begin{figure}[ptb]
\begin{center}
%\vspace{-3cm} \hspace{-0.93cm}
\includegraphics[scale=0.42]{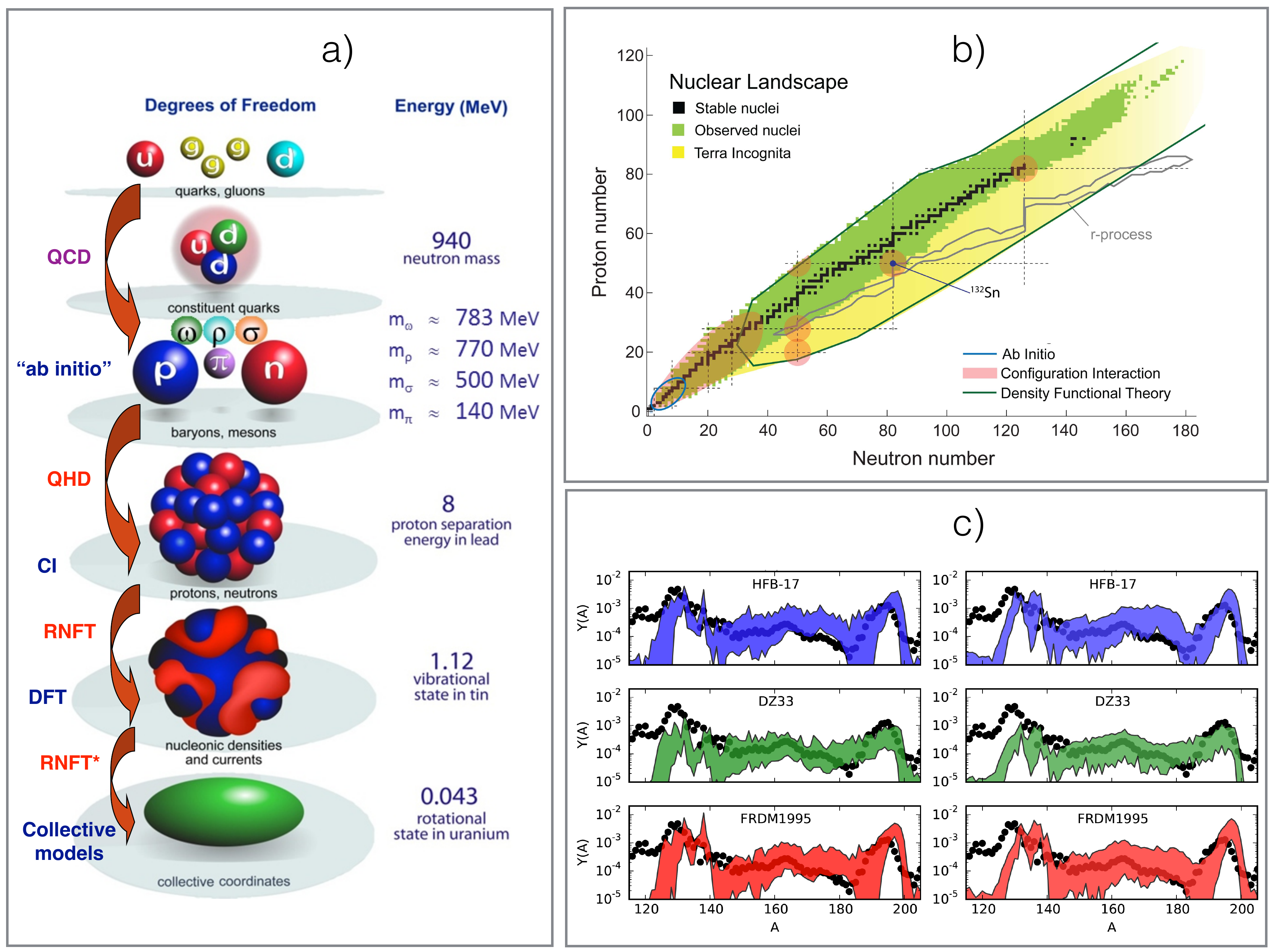}
\vspace{-1cm}
\end{center}
\caption{
(a) Characteristic resolution scales and the corresponding degrees of freedom in nuclear physics (adopted from Ref. \cite{BertschDeanNazarewicz2007} ). The traditional theoretical nuclear structure concepts associated with the given scales are indicated as 
"ab initio", Configuration Interaction (CI), and Density Functional Theory (DFT). The relativistic theories, constrained by the  low-energy 
Quantum Chromodynamics (QCD) and considered in the present work, are indicated by the red font color: Quantum Hadrodynamics (QHD), Relativistic Nuclear Field Theory (RNFT), and an extension to the rotational degrees of freedom as RNFT*.  (b) Chart of the nuclei adopted from Refs. \cite{Bertsch2007,LitvinovaBrownFangEtAl2014}, representing the domains of the major nuclear structure concepts.
(c) Variance in isotopic abundance patterns from uncertain beta decay half-lives (left) and uncertain neutron
capture rates (right) for the most commonly used phenomenological mass models, adopted from Ref. \cite{MumpowerSurmanMcLaughlinEtAl2016}.} 
\label{scales}%
\vspace{-0.2 cm}
\end{figure}
%

%However, besides their limited applicability, the microscopic theoretical methods described above have conceptual limitations: 
%(i) The most advanced "ab initio" chiral effective field theory ($\chi$EFT) is sensitive to the momentum cutoff and has inherent problems associated with the presence %of the contact term \cite{Machleidt2015};
%(ii) The shell-model describes nuclear many-body correlations very well, but, except for the no-core shell-model, which is rather classified as "ab initio", it is based on %the concept of the inert core and valence space and uses an effective nucleon-nucleon interaction, which is often disconnected from the shell structure. (iii) Self-%consistent density functional theories do not allow for a high-precision description of nuclear properties due to a very poor treatment of many-body correlations, which %are especially important for exotic systems at the extremes of nuclear matter. Delicate interplay of various kinds of correlations is responsible for binding of loosely 5bound systems, their shapes, decay properties, and for low-energy spectra.

As the physics at a given energy scale is governed by the next ("underlying") higher energy scale, nuclear many-body theories naturally tend to be as fundamental as possible to increase their predictive power. Chronologically, the general vector of their development points to resolving more and more fundamental scales and, as mentioned above, a successful description of light nuclei is already possible in LQCD, due to the powerful computer technologies and advanced numerical algorithms. However, even if LQCD succeeds in computing heavy nuclear systems, such a computation "would give one no understanding of the physical nature of the nuclear phenomena" \cite{MigdalSapersteinTroitskyEtAl1990}, because understanding implies a consistent scale-connecting theory. 

Fig. \ref{scales} (a) adopted from Ref. \cite{BertschDeanNazarewicz2007} shows schematically the characteristic resolution scales in nuclear physics and the associated degrees of freedom. One can see that between the typical proton separation energy and the pion mass there is a gap of about one order of magnitude, which justifies a considerable success of the pionless theories based on the nucleonic degrees of freedom in the energy region below ~50 MeV, commonly associated with the nuclear structure domain. 
A similar, although less clear separation, exists between one-nucleon and vibrational scales. However, this energy gap practically disappears in loosely bound nuclear systems and, indeed, it is known since A. Bohr and B.R. Mottelson \cite{BohrMottelson1969,BohrMottelson1975} that nucleonic and vibrational degrees of freedom are strongly coupled at least in medium-mass and heavy nuclei. Finally, the gap between the typical vibrational and rotational degrees of freedom is formally of 1-2 orders of magnitude, however, there is a large variety of nuclei of transitional character, where vibrations and rotations are essentially coupled. Thus, since nature does not allow for a clear scale separation, a successful theory should be able to connect them. 
Fig. \ref{scales}(b) shows the approximately outlined domains of the main theoretical concepts on the nuclear landscape. Predictive "ab initio" theories are capable of describing the properties of light nuclei, although their reach is currently extended to the calcium mass region. Shell-models perform well in the areas colored by pink: up to $Z \simeq N \simeq$ 40 and around the shell closures. The DFT has the largest domain covering almost the entire nuclear chart except for the lightest nuclei. A remarkable feature of this picture is that different models can constrain each other in the overlapping domains, as discussed in Ref. \cite{LitvinovaBrownFangEtAl2014}. 
Fig. \ref{scales}(c) adopted from Ref. \cite{MumpowerSurmanMcLaughlinEtAl2016} illustrates the uncertainties in the isotopic abundance pattern due to the uncertainties in the rates of the two phases of r-process nucleosynthesis - radiative neutron capture and beta decay. The rates are given by the most commonly used phenomenological mass models with the corresponding theoretical uncertainties. The degree of error propagation is clearly very high, that indicates the importance of a reliable and fully microscopic underlying nuclear structure theory. 

As a possible solution we developed throughout  the last decade a relativistic nuclear field theory (RNFT) \cite{LitvinovaRing2006,LitvinovaRingTselyaev2007,LitvinovaRingTselyaev2008,LitvinovaRingTselyaev2010,Litvinova2015,Litvinova2016} built on Quantum Hadrodynamics (QHD), which is, in turn, constrained by the low-energy QCD. QHD, being a covariant theory of interacting nucleons and mesons, turned out to be very successful on the mean field level \cite{BogutaBodmer1977,SerotWalecka1979,SerotWalecka1986a,SerotWalecka1997,Ring1996}. The idea of fine-tuning meson masses and coupling constants together with introducing a non-linear scalar meson lead to a very good quantitative description of nuclear ground states. Thus, QHD has provided the connection between the low-energy QCD scale and the nucleonic scale in the complex nuclear medium. The time-dependent versions of the relativistic mean field (RMF) model and the response theory built on it have allowed a very good description of the positions of collective vibrational states in the relativistic random phase approximation (RRPA)\cite{RingMaVanGiaiEtAl2001,VretenarAfanasjevLalazissisEtAl2005,LiangVanGiaiMengEtAl2008}  or, for the superfluid systems, by the quasiparticle RRPA (RQRPA) \cite{PaarRingNiksicEtAl2003}. The RMF and R(Q)RPA form the content of the covariant DFT (CDFT), which performs amazingly well and provides a description of nuclear properties, comparable to that of other DFT's which are not based on the Lorentz symmetry and meson-exchange interaction and have a larger number of adjustable parameters.%
\begin{figure}
\begin{center}
%\vspace{-0.5cm}
\includegraphics[scale=0.35]{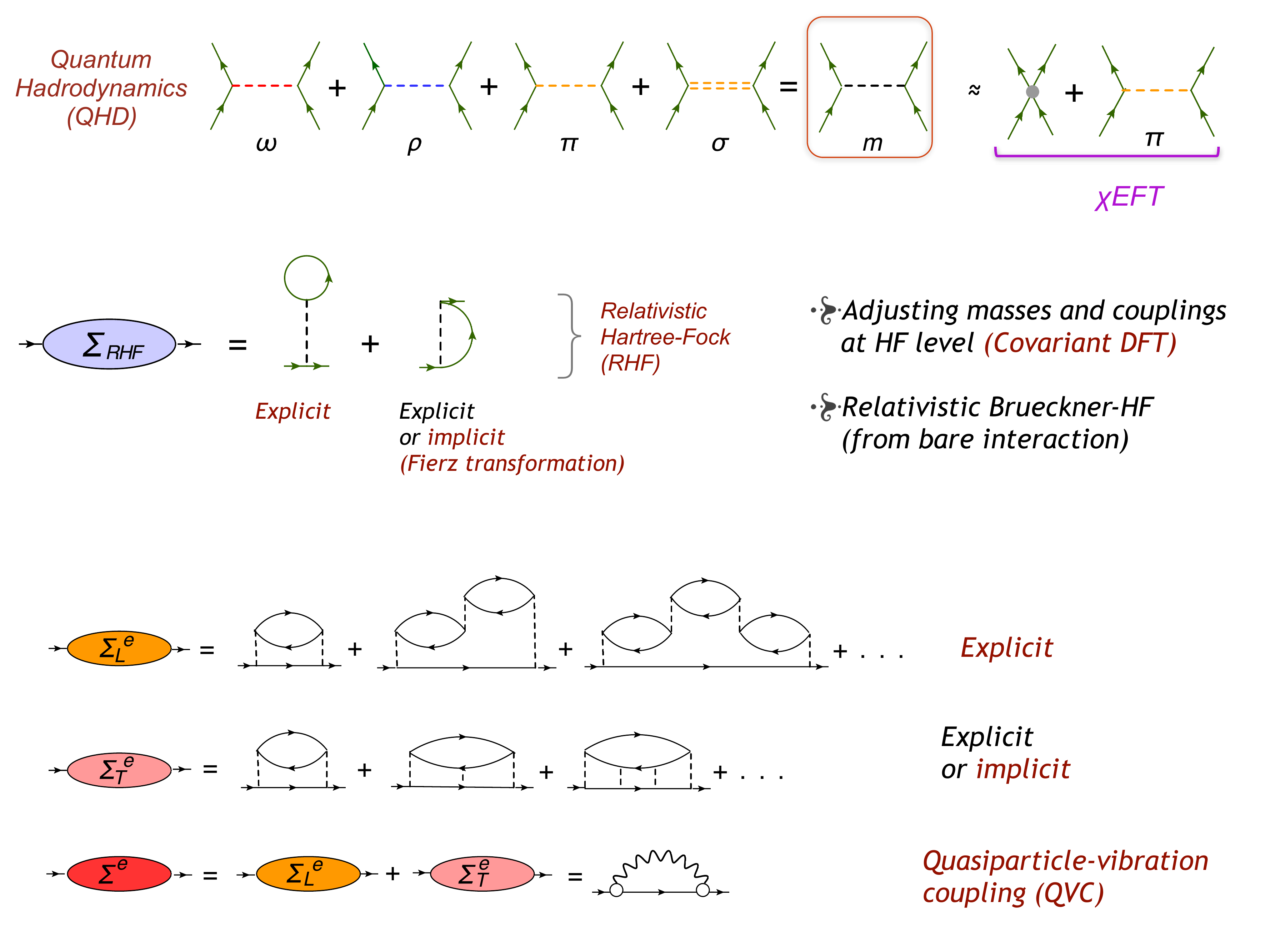}
\end{center}
\vspace{-0.5cm}
\caption{The meson-exchange interaction in the diagrammatic form (top); the nucleonic self-energy in the relativistic Hartree-Fock approximation (middle); the random phase approximation to the time-dependent (energy-dependent) part of the nucleonic self-energy (bottom).}
\label{sfe}%
\end{figure}

However, even CDFT lacks the above mentioned many-body correlations, first of all, those associated with retardation and temporal non-localities of the nucleonic self-energy and effective interaction. Therefore, the next step connecting single-nucleon and vibrational scales was made by the RNFT \cite{LitvinovaRing2006,LitvinovaRingTselyaev2007,LitvinovaRingTselyaev2008,LitvinovaRingTselyaev2010,LitvinovaRingTselyaev2013,Litvinova2015,Litvinova2016}, a relativistic version of the original NFT \cite{BesBrogliaDusselEtAl1976,BortignonBrogliaBesEtAl1977,BertschBortignonBroglia1983,MahauxBortignonBrogliaEtAl1985,ColoBortignon2001,NiuColoVigezziEtAl2014,NiuNiuColoEtAl2015},
which accounts for retardation effects of meson exchange, missing in CDFT, in an approximate way. The approximation is based on the emergence of collective degrees of freedom, such as vibrations (phonons) caused by coherent nucleonic oscillations, as shown in Fig. \ref{sfe}. An order parameter associated with (quasi)particle-vibration coupling (QVC) vertices provides a consistent power counting with respect to this parameter and controlled truncation schemes. The great advantage of the RNFT is that the QVC vertices, which give the most important contributions to the nucleonic self-energy and effective interaction beyond CDFT, can be well approximated by an infinite sum of the ring diagrams of RRPA or, for the superfluid systems, of RQRPA. The non-perturbative treatment of the QVC effects, which are responsible for the retardation in RNFT, is based on the time ordering of the two-loop and higher-order diagrams, containing multiple exchange of vibrations between nucleons, and on the evaluation of their relative contributions to the one- and two-nucleon propagators. The truncation schemes justified by this evaluation form the content of the so-called time blocking approximation which has been proposed by V.I. Tselyaev for up to two-loop diagram contributions (or 2p-2h type, namely, 2q$\otimes$phonon configurations) to the effective interaction \cite{Tselyaev1989}, adopted to the relativistic framework in Ref. \cite{LitvinovaRingTselyaev2008} and generalized recently  for an arbitrary order of complexity (np-nh, or 2q$\otimes$Nphonon) \cite{Litvinova2015}. Lately, pion-nucleon correlations beyond the Hartree-Fock approximation have been included in RNFT in the form of exchange by isospin-flip phonons \cite{Litvinova2016}.

The nuclear response theory with QVC, which we developed within this formalism in a parame-ter-free way and called relativistic quasiparticle time blocking approximation (RQTBA), has provided a high-quality description of gross properties of the giant resonances \cite{LitvinovaRingTselyaev2008,MarketinLitvinovaVretenarEtAl2012,LitvinovaBrownFangEtAl2014} and some fine features of excitation spectra at low energies \cite{EndresLitvinovaSavranEtAl2010,RobinLitvinova2016} in both neutral and charge-exchange channels for medium-mass and heavy nuclei.  In particular, the isospin splitting of the pygmy dipole resonance has been explained quantitatively \cite{EndresLitvinovaSavranEtAl2010} and the beta-decay half-lives were reproduced very successfully in the first version of the proton-neutron RQTBA (pn-RQTBA) \cite{RobinLitvinova2016}. The recently proposed generalized RQTBA with multiphonon couplings \cite{Litvinova2015} opens a way to unify the theory of high-frequency collective oscillations and low-energy spectroscopy.
The important features of the RNFT are that (i) it is constrained by the fundamental underlying theory, such as QCD, and hence, consistent with Lorentz invariance, parity invariance, electromagnetic gauge invariance, isospin and chiral symmetry (spontaneously broken) of QCD; (ii) it connects the scales from the low-energy QCD degrees of freedom to complex collective phenomena self-consistently, i.e. without introducing new parameters throughout the connection;
(iii) it includes effects of nuclear superfluidity on equal footing with the meson exchange and QVC, so that it is applicable to open-shell nuclei; (iv) it is applicable and demonstrates a high quality of performance throughout almost the entire nuclear chart, from the oxygen mass region to superheavy nuclei \cite{LitvinovaAfanasjev2011,Litvinova2012}.

In Refs. \cite{LitvinovaRing2006,LitvinovaAfanasjev2011,AfanasjevLitvinova2015,Litvinova2012} as well as in the applications to the nuclear response \cite{LitvinovaRingTselyaev2007,LitvinovaRingTselyaev2008,LitvinovaRingTselyaev2010,EndresLitvinovaSavranEtAl2010,LitvinovaBrownFangEtAl2014,RobinLitvinova2016}, we have included coupling between (quasi)particles and isoscalar phonons of natural parities, as it is done traditionally in the NFT, into the nucleonic self-energy. As soon as a consistent description of the isospin-flip excitations has become available \cite{MarketinLitvinovaVretenarEtAl2012,LitvinovaBrownFangEtAl2014,RobinLitvinova2016}, in the more recent implementations of RNFT we include also isospin-flip phonons in the one-nucleon self-energy.
As in Refs. \cite{LitvinovaRing2006,LitvinovaAfanasjev2011,AfanasjevLitvinova2015,Litvinova2012} and many other applications, we continue to use the diagonal approximation for the proper self-energy.
In a spherical system without superfluid pairing, the proper neutron and proton self-energies $\Sigma^{e}_{(n)}(\varepsilon)$ and $\Sigma^{e}_{(p)}(\varepsilon)$, including both non-isospin-flip and isospin-flip phonons, have the following form:
\begin{eqnarray}
\Sigma^{e}_{(n)}(\varepsilon) = \frac{1}{2j_{n}+1}\Bigl[ \sum\limits_{(\mu n^{\prime})} \frac{|\gamma_{(\mu;nn^{\prime})}^{\eta_{n\prime};\eta_n\eta_{n\prime}} |^2}{\varepsilon - \varepsilon_{n\prime} - \eta_{n\prime}(\Omega_{\mu} - i\delta)} + 
\sum\limits_{(\lambda p^{\prime})} \frac{|\zeta_{(\lambda;np^{\prime})}^{\eta_{p\prime};\eta_n\eta_{p\prime}} |^2}{\varepsilon - \varepsilon_{p\prime} - \eta_{p\prime}(\omega_{\lambda} - i\delta)}\Bigr], \label{se1a}\\
\Sigma^{e}_{(p)}(\varepsilon) = \frac{1}{2j_{p}+1}\Bigl[\sum\limits_{(\mu p^{\prime})} \frac{|\gamma_{(\mu;pp^{\prime})}^{\eta_{p\prime};\eta_p\eta_{p\prime}}  |^2}{\varepsilon - \varepsilon_{p\prime} - \eta_{p\prime}(\Omega_{\mu} - i\delta)} +
\sum\limits_{(\lambda n^{\prime})} \frac{| \zeta_{(\lambda;pn^{\prime})}^{\eta_{n\prime};\eta_p\eta_{n\prime}}  |^2}{\varepsilon - \varepsilon_{n\prime} - \eta_{n\prime}(\omega_{\lambda} - i\delta)}\Bigr], 
\label{se1b}
\end{eqnarray}
where 
\begin{eqnarray}
\gamma_{(\mu;nn^{\prime})}^{\eta_{\mu};\eta_n\eta_{n\prime}} =   \gamma_{(\mu;nn^{\prime})}^{\eta_n\eta_{n^{\prime}}}  \delta_{\eta_{\mu},+1} + \gamma_{(\mu;n^{\prime} n)}^{\eta_{n^{\prime}}\eta_n}  \delta_{\eta_{\mu},-1}, \ \ \ \ \ \ \ \ \ \ \gamma_{(\mu;nn^{\prime})}^{\eta_n\eta_{n^{\prime}}}  = \langle n \parallel \gamma_{(\mu)}^{\eta_n\eta_{n^{\prime}}} \parallel n^{\prime}\rangle \nonumber \\
\zeta_{(\lambda;np^{\prime})}^{\eta_{\lambda};\eta_n\eta_{p\prime}} =  \zeta_{(\lambda;np^{\prime})}^{\eta_n\eta_{p^{\prime}}} \delta_{\eta_{\lambda},+1} + \zeta_{(\lambda;p^{\prime} n)}^{\eta_{p^{\prime}}\eta_n} \delta_{\eta_{\lambda},-1}\ \ \ \ \ \ \ \ \ \ \zeta_{(\lambda;np^{\prime})}^{\eta_n\eta_{p^{\prime}}}  = \langle n \parallel \zeta_{(\lambda)}^{\eta_n\eta_{p^{\prime}}} \parallel p^{\prime}\rangle.
\label{gammazeta}
\end{eqnarray}
%and
%\begin{eqnarray}
%\gamma_{(\mu;nn^{\prime})}^{\eta_n\eta_{n^{\prime}}}  = \langle n \parallel \gamma_{(\mu)}^{\eta_n\eta_{n^{\prime}}} \parallel n^{\prime}\rangle \nonumber \\
%\zeta_{(\lambda;np^{\prime})}^{\eta_n\eta_{p^{\prime}}}  = \langle n \parallel \zeta_{(\lambda)}^{\eta_n\eta_{p^{\prime}}} \parallel p^{\prime}\rangle,
%\label{gammazeta}
%\end{eqnarray}
Here $\eta_k = \pm 1$ for the particle (hole) states and $\varepsilon_k$ are the mean-field single-particle energies.  In Eqs. (\ref{se1a}), (\ref{se1b}) the coupling to the isoscalar phonons is described by the vertices $\gamma_{\mu}$ and frequencies $\Omega_{\mu}$, and to isovector (proton-neutron (pn) and neutron-proton (np)) ones by the vertices $\zeta_{\lambda}$ and frequencies $\omega_{\lambda}$.  
%
%\begin{figure*}
%\begin{center}
%\vspace{-6cm}
%\includegraphics[scale=0.7]{Self-energies1.eps}
%\end{center}
%\caption{Isospin structure of the proper self-energy in the second-order of PVC.  }
%\label{figse}%
%\end{figure*}
%
These quantities are extracted from the response function (in the RRPA approximation in the leading order) of the corresponding multipolarities $J_{\mu}$, $J_{\lambda}$. For the isoscalar phonons of particle-hole nature this procedure is described in detail in Refs. \cite{LitvinovaRing2006,LitvinovaRingTselyaev2008}. First we run RRPA calculations for the particle-hole (ph) and hole-particle (hp) components of the vertices $\gamma_{\mu}$ and then determine the particle-particle (pp) and hole-hole (hh) components - all of them enter the first sums of the Eqs. (\ref{se1a}), (\ref{se1b}).  The characteristics of the isovector phonons are computed analogously using the approach of Ref. \cite{MarketinLitvinovaVretenarEtAl2012} on the level of the proton-neutron RRPA (pn-RRPA).
They determine the residues and poles of the second sums in Eqs. (\ref{se1a}), (\ref{se1b}). The indices in the round brackets in these equations indicate that magnetic quantum numbers are excluded (reduced matrix elements), and the analogous convention for the proton-proton and proton-neutron matrix elements applies.
\begin{figure}
\begin{center}
%\vspace{-0.5cm}
\includegraphics[scale=0.425]{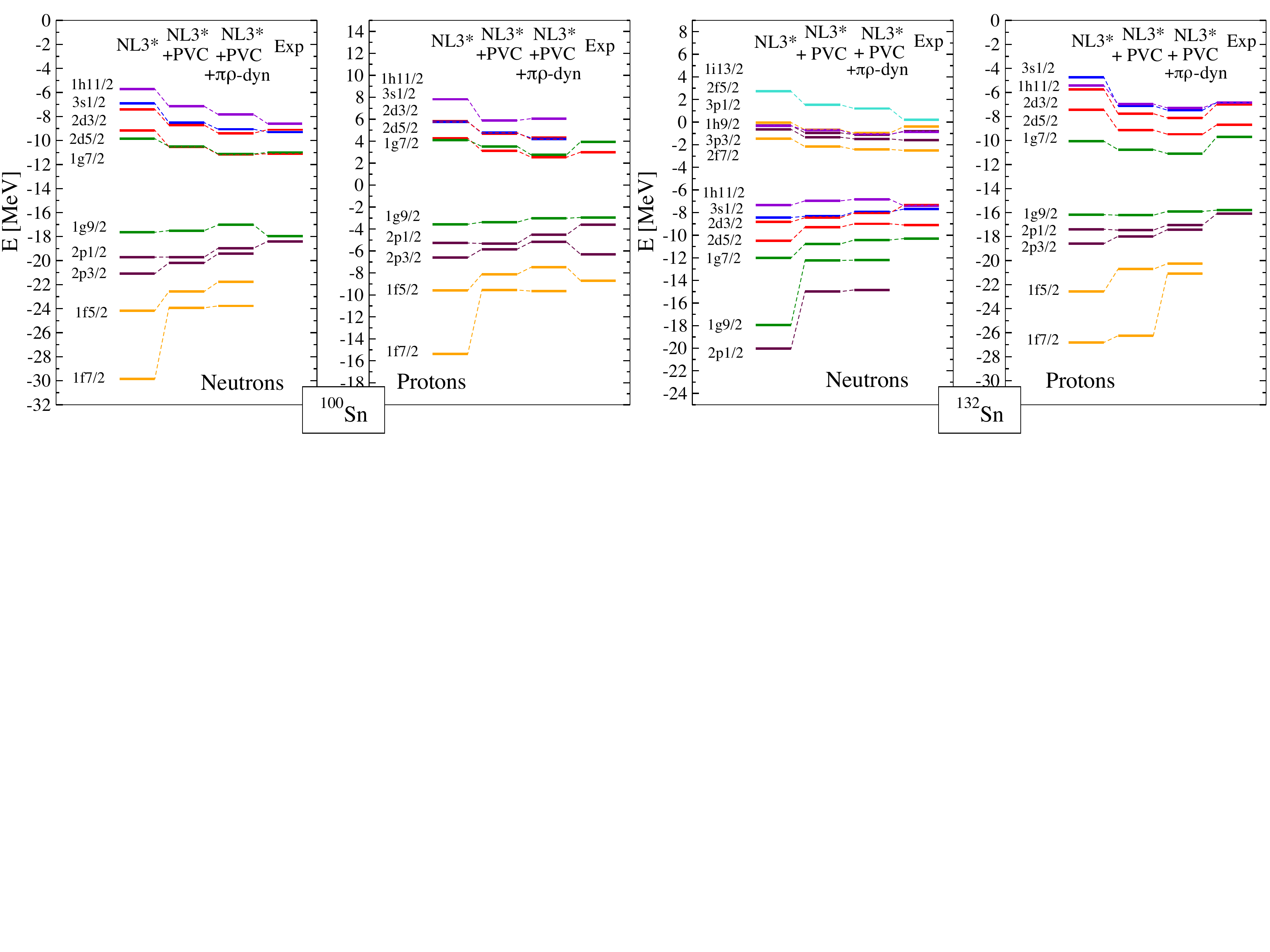}
\end{center}
\vspace{-6.7cm}
\caption{Single-particle states in $^{100,132}$Sn calculated in the RMF (NL3*) and in extended approaches with particle-vibration coupling (PVC) \cite{Litvinova2016}.}
\label{spstates}%
\end{figure}

After that, the Dyson equation formulated in \cite{LitvinovaRing2006} is solved with the self-energy (\ref{se1a}), (\ref{se1b}), and the energies of the fragmented single-particle levels together with the corresponding spectroscopic factors are obtained. The ground state correlations associated with $\pi$- and $\rho$-meson dynamics and expressed by the terms with $\eta_{p'} \neq \eta_{n}$ and $\eta_{n'} \neq \eta_{p}$  are neglected in the calculations. Further details can be found in Ref. \cite{Litvinova2016}.
The results of these calculations with NL3* parameter set \cite{LalazissisKaratzikosFossionEtAl2009} for the single-particle states in $^{100,132}$Sn are displayed in Fig. \ref{spstates}. The mean-field states are shown in the first columns from left (NL3*) of each panel and the (NL3*+PVC) calculations with the self-energy including only isoscalar phonons are given in the second columns. The third columns (NL3*+PVC+$\pi\rho$-dyn) display the results obtained with the additional contribution of the isospin-flip phonons, and the fourth columns (Exp) present the 'experimental' single-particle energies  extrapolated from data \cite{GBST.14,GLM.07}. The two middle columns contain only the dominant single-particle states, i.e. those with the maximal spectroscopic amplitudes.  

In Ref. \cite{LitvinovaAfanasjev2011}, the influence of the isoscalar phonons on the single-particle spectra was investigated systematically within the PVC model, which has demonstrated a significant overall improvement of the description of the dominant single-particle states. This approach is shown in columns (NL3*+PVC) whose difference with the next columns (NL3*+PVC+$\pi\rho$-dyn) reveals the dynamical contribution of $\pi$ and $\rho$-mesons to the positions of the dominant single-particle states. 
%The contributions of the isospin-flip phonons are associated with mainly pionic processes, because the analysis of the matrix elements of the interaction (${\tilde V}_%\rho + {\tilde V}_\pi + {\tilde V}_{\delta \pi}$) in Ref. \cite{LZRRM.12} shows that the contribution from the $\rho$-meson exchange is relatively small. 
% Main results - modified/extended according the referee's request
As can be seen from Fig. \ref{spstates}, the inclusion of the $\pi$- and $\rho$-meson dynamics provides additional shifts of the dominant single-particle states. These shifts amount from a few hundreds keV to 1 MeV. Only little changes are found in the spectroscopic factors for the major part of the considered states, as compared to Ref. \cite{LitvinovaAfanasjev2011}. Overall, the impact of the isovector phonons on the dominant states is weaker than that of the isoscalar ones, however, for some states far from the Fermi surface like, for instance, the proton state 1f$_{7/2}$ in $^{132}$Sn, the effect is significant, because of the change of the dominant fragment due to the redistribution of the strength. Further inclusion of the dynamical $\pi$- and $\rho$-meson ground state correlations may introduce some minor changes in the picture shown in Fig. \ref{spstates}, which will be investigated in the future. 

\begin{figure}
\begin{center}
%\vspace{-1.0cm}
\includegraphics[scale=0.42]{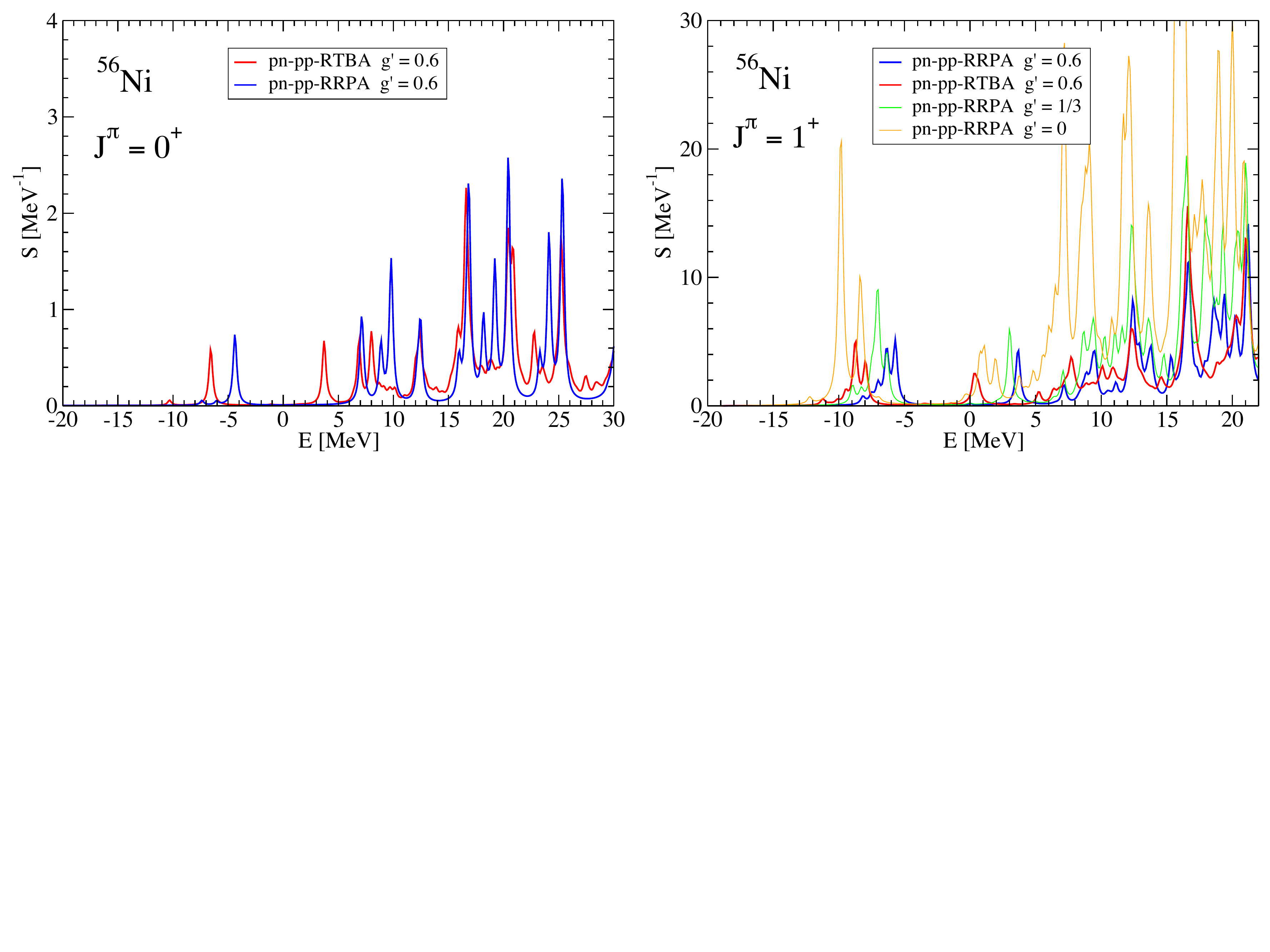}
\end{center}
\vspace{-6cm}
\caption{Deuteron-addition strength distribution in $^{56}$Ni calculated in pn-pp-RRPA and pn-pp-RTBA.}
\label{pn-pp-rtba}%
\end{figure}
As it is known from the original NFT \cite{BesBrogliaDusselEtAl1976,BortignonBrogliaBesEtAl1977,BertschBortignonBroglia1983}, pairing vibrations should also play a noticeable role in one-nucleon self-energy, however, so far they were not investigated in the relativistic NFT framework. As a first step to fill this gap, we have recently started to build an approach to the isospin-flip, or proton-neutron, pairing vibrations. As an illustration, in Fig. \ref{pn-pp-rtba} we show preliminary results for such vibrations in the form of response to the deuteron addition operators in J$^{\pi}=0^+$ and J$^{\pi}=1^+$ channels. The calculations within the proton-neutron particle-particle RRPA (pn-pp-RRPA) and within the proton-neutron particle-particle RTBA (pn-pp-RTBA) are shown to reveal the role of the retardation effects of PVC on these modes. The calculations are based on the NL3 meson-exchange interaction \cite{LalazissisKonigRing1997} and free-space pion-nucleon coupling with the zero-range Landau-Migdal term, see Ref. \cite{LitvinovaRobinEgorova2016} for more details. The pn-pp-RRPA results with various values of the strength of the repulsive Landau-Migdal interaction are shown by blue, green and orange curves. The right panel of  Fig. \ref{pn-pp-rtba} displays the evolution of the pn-pp-RRPA strength with the change of this parameter from its realistic value $g'$ = 0.6 to its complete disappearance, in order to see the sensitivity of the response in the pn-pp channel to this parameter. The red curves give the pn-pp-RTBA strength distributions with $g'$ = 0.6 illustrating the effect of PVC induced by the coupling to the isoscalar phonons of natural parities. The latter effect leads to some fragmentation of the strength and its redistribution to lower energies. A more detailed study of these excitation modes and of their contribution to the nucleonic self-energy will be given elsewhere.

In conclusion, we have discussed some recent advancements of RNFT which allow us (i) to include the effects of isovector $\pi$- and $\rho$-mesons beyond the Hartree(-Fock) approximation in the theory and (ii) to investigate proton-neutron pair transfer excitations. The former is found rather significant for the nuclear shell structure, at least, at the present level of description and the latter opens the way to understanding effects of proton-neutron pairing: its underlying mechanisms and its influence on various nuclear structure observables.

\end{document}